\documentclass[sigconf]{aamas} 
\usepackage{balance} % for balancing columns on the final page
\usepackage{xcolor}
\usepackage{todonotes}
\usepackage{dsfont}
\usepackage{bm}

\setcopyright{ifaamas}
\acmConference[AAMAS '23]{Proc.\@ of the 22nd International Conference
on Autonomous Agents and Multiagent Systems (AAMAS 2023)}{May 29 -- June 2, 2023}
{London, United Kingdom}{A.~Ricci, W.~Yeoh, N.~Agmon, B.~An (eds.)}
\copyrightyear{2023}
\acmYear{2023}
\acmDOI{}
\acmPrice{}
\acmISBN{}

\acmSubmissionID{1108}

\title[Decentralized core-periphery structure accelerates innovation in ABMs]{Decentralized core-periphery structure in social networks accelerates cultural innovation in agent-based model}
\author{Jesse Milzman}
\affiliation{
\institution{DEVCOM Army Research Laboratory}
\city{Adelphi, MD}
\country{United States}}
\email{jesse.m.milzman.civ@army.mil}

\author{Cody Moser}
\affiliation{
\institution{University of California Merced}
\city{Merced, CA}
\country{United States}}
\email{cmoser2@ucmerced.edu}

\begin{abstract}
Previous investigations into creative and innovation networks have suggested that intellectual, artistic, and technological innovations often occur with the interaction of core and peripheral actors, with both individuals and teams at intermediary positions well-situated to innovate.
In this work, we investigate the effect of global core-periphery network structure on the speed and quality of cultural innovation.
Drawing on differing notions of core-periphery structure from \cite{priebe2019two} and \cite{borgatti2000models}, we distinguish decentralized core-periphery, centralized core-periphery, and affinity network structure.
We generate networks of these three classes from stochastic block models (SBMs), and use them to run an agent-based model (ABM) of collective cultural innovation, in which agents can only directly interact with their network neighbors.
In order to discover the highest-scoring innovation, agents must discover and combine the highest innovations from two completely parallel technology trees.
We find that decentralized core-periphery networks outperform both centralized core-periphery networks and affinity networks, in terms of mean crossover time for this final innovation.
We hypothesize that decentralized core-periphery network structure provides a more fruitful environment for collective problem-solving, by allowing for the relative shielding of periphery nodes from the optimal innovations known by the core community at any given time.
This prevents the disincentive for parallel explorations that emerges in a highly connected, centralized network.
We then build upon the "Two Truths" hypothesis regarding community structure in spectral graph embeddings first articulated in \cite{priebe2019two}.
The Two Truths hypothesis suggests that the adjacency spectral embedding (ASE) captures core-periphery community structure in a graph, while the laplacian spectral embedding (LSE) captures affinity structure.
For a given network generated from an SBM, we use ASE and LSE to resample new networks of similar structure, using either to parametrize a random dot product graph (RDPG) model.
We find that, for core-periphery networks, ASE resampling best recreates networks with similar performance on the innovation SBM.
Since the Two Truths hypothesis suggests that ASE captures core-periphery structure, this result further supports our hypothesis.
\end{abstract}


\newcommand{\BibTeX}{\rm B\kern-.05em{\sc i\kern-.025em b}\kern-.08em\TeX}

\DeclareMathOperator{\SBM}{SBM}
\DeclareMathOperator{\ER}{ER}
\DeclareMathOperator{\ASERDPG}{\text{ASE-RDPG}}
\DeclareMathOperator{\LSERDPG}{\text{LSE-RDPG}}

\newcommand{\etal}[1]{\textit{et al.~}}
\begin{document}

%%% The following commands remove the headers in your paper. For final 
%%% papers, these will be inserted during the pagination process.

\pagestyle{fancy}
\fancyhead{}

%%% The next command prints the information defined in the preamble.

\maketitle 

%%%%%%%%%%%%%%%%%%%%%%%%%%%%%%%%%%%%%%%%%%%%%%%%%%%%%%%%%%%%%%%%%%%%%%%%

\section{Introduction}
From a broad perspective, innovation is understood as a form of collective problem-solving. For this and other reasons, the process of innovation is understood as a social process, as social collectives are capable of, and in some cases optimized for, both retaining the knowledge of previous generations while building upon this knowledge for subsequent innovations, a phenemonon we refer to as "cumulative" culture \cite{migliano2017characterization,migliano2020hunter}. Human social networks tend to exhibit core-periphery structures, whereby a `core' population is heavily inter-connected, and connected in turn to more `peripheral' individuals and subcommunities \cite{borgatti2000models}.
Prior work on the structure of human networks has suggested that innovation emerges at the boundary between the core and periphery of creative networks \cite{cattani2008core,centola2021influencers}.
Individual innovators are often in an intermediate position with many core and peripheral connections, and successfully innovative teams tend to include both core and peripheral individuals \cite{cattani2008core}.

% \begin{itemize}
%   \color{red}
%   \item relationshp of different latent space embeddings to core-periphery structure
% \end{itemize}

However, to our knowledge, there has not been an in-depth examination of the relationship between the \textit{global} core-periphery structure of a creative population and its innovation potential.
Moreover, there are multiple candidate metrics purporting to quantify core-periphery structure within a given network, suggesting that the concept is ill-posed \cite{rombach2014core}.
Most definitions of core-periphery structure, formal and informal, assume that core nodes ought to be be heavily connected to peripheral nodes \cite{borgatti2000models,rombach2014core}.
Nonetheless, at least one recent work suggests a definition that precludes this feature \cite{priebe2019two}, as it is not observed in the core-periphery structure observed in connectome networks \cite{priebe2019two}, corresponding to the grey/white matter divide.
The authors of \cite{priebe2019two} instead define core-periphery networks such that peripheral nodes are only loosely connected to either core or periphery, with comparable probabilities for the edges to either type.
Thus, even in the simplest setting in which one ignores the possibility of multiple cores or intermediate layers, there is a conceptual distinction that must be made between the desired expected relationship between core and peripheral nodes.

We therefore utilize an agent-based model (ABM) of collective problem-solving known as the "Potions Game" in order to investigate the effects of core-periphery structure on the ability of a population of agents to locate optimal solutions in an innovation task. In the model, agents are given a set of ingredients which can be combined with the ingredients of one of their neighbors to produce new ones. These novelties are shared with their neighbors and can be combined with the previous set of ingredients to produce additional novelties. As new ingredients are discovered, agents are more likely to use them over old ones, but unbeknownst to the agents, the initial set of ingredients can be combined to produce new ingredients along two separate pathways. Because agents are unlikely to return to the original set of ingredients, this creates path dependency in the model, but if agents are able to find and exploit both pathways, they can combine their final ingredient to discover an ingredient which successfully ends the simulation.

By running our ABM on core-periphery networks, we may explore the distinctions between network structures along several dimensions, and their implications for collective problem-solving. As already mentioned, \cite{borgatti2000models} and \cite{priebe2019two} suggest distinct notions of core-periphery structure, reflecting the degree to which the network core dominates the peripheral neighborhood. Thus, we split core-periphery structure into the two types suggested: one with a periphery which is not very strongly connected to itself, but is strongly connected to the core (a centralized periphery); and one with a periphery which is isolated both from itself and from the core (a decentralized periphery).
Moreover, core-periphery networks in \cite{priebe2019two} are contrasted to ``affinity''  networks, in which both communities have more within than between similarity.

% In addition to running an agent-based model on core-periphery networks, we explore the phenomenon of core-periphery structure along several dimensions. Prior work in spectral graph clustering has noted where two separate methods for identifying structure will identify separate clusters roughly representing "affinity" networks and representing "core-periphery" networks whereby affinity networks are those where communities have more within than between similarity and core-periphery networks are those with a similar core and less-connected periphery. In a recent study exploring these two types of clusters, researchers found that core-periphery structure itself could be split into two dimensions: one with a periphery which is not very strongly connected to itself, but is strongly connected to the core (a centralized periphery); and one with a periphery which is isolated both from itself and from the core (a decentralized periphery). We hypothesize that in centralized core-periphery structures, peripheral nodes conform to the behavior of the core and therefore limit diversity in the network while in decentralized core-periphery structures, peripheral nodes are able to freely "explore" possible alternative combinations in the Potions Game while the core exploits whichever innovation pathway they are on.

Finally, we discover a surprising crossover relationship between decentralized core-periphery structure, spectral graph embeddings, and performance on our innovation ABM.
The adjacency spectral embedding (ASE) \cite{sussman2012consistent} and the Laplacian spectral embedding (LSE) \cite{von2007tutorial} are common methods to embed graphs in Euclidean space for the purpose of community detection \cite{von2007tutorial,sussman2012consistent,lyzinski2016community}.
However, despite statistical results suggesting both are consistent estimators for theoretical models, they do not always recover the same communities. Priebe \textit{et al.} observed that ASE seems to recover core-periphery communities, while LSE recovers affinity communities \cite{priebe2019two}.
Keeping in mind this difference, we investigate the use of ASE and LSE to resample networks with a given structure (core-periphery or affinity), and compare the ABM performance of the resampled networks to the original.

% In line with the results from \cite{priebe2019two}, we find that ASE-based resampling better recreates ABM performance for core-periphery networks, compared to the LSE-based method.

% There is a rich literature on the use of spectral decompositions to embed graphs in Euclidean space.\todo{Introduce ASE and LSE in language, not form}
% In particular, spectral clustering, the combination of spectral embedding and standard clustering methods in Euclidean space (K-means, GMMs), form a popular and ubiquitous class of methods used in data science for handling network and/or high-dimensional data sets.
% Central limit theorems exist suggesting that both ASE and LSE provide asymptotically consistent estimates of the latent positions of random dot product graphs (RDPGs), of which SBMs are a special case \cite{sussman2013consistent,tang2018limit}.

%%%%%%%%%%%%%%%%%%%%%%%%%%%%%%%%%%%%%%%%%%%%%%%%%%%%%%%%%%%%%%%%%%%%%%%%

\section{Related Work}
Network scientists have collectively formulated a vast taxonomy of structural qualities to describe network topologies, including those of density/sparsity, efficiency, hierarchy \cite{mengistu2016evolutionary,lyzinski2016community}, affinity/modularity \cite{newman2006modularity,clune2013evolutionary}, and core-periphery separation \cite{borgatti2000models}.
This last quality has been implicated in both individual and collective learning, via the analysis of creative and learning networks, respectively \cite{centola2022network,cattani2008core,cattani2015creativity,bassett2013task}.

One early attempt to formalize a model of core-periphery structure was that of Borgatti and Everett \cite{borgatti2000models}.
Grounding themselves in block modeling approaches common in social network science at the time, they identified ideal core-periphery networks as two-block graphs in which the nodes of one block (the `core') are fully connected to every other node, while the nodes in the other block (`periphery') are connected only to the core nodes.
They propose a statistical test for an \textit{a priori} core-periphery partition, whose statistic is the Pearson coefficient between the entries of the adjacency matrix and those of the ideal block model for the same partition and also propose an algorithm for assigning every node a continuous coreness value for weighted networks.
Since \cite{borgatti2000models}, multiple works have developed new metrics \cite{rombach2014core,rombach2017core,tudisco2019nonlinear,gallagher2021clarified,yanchenko2022core}, both for classifying nodes according to coreness, and detecting more complex structure such as multiple cores \cite{kojaku2017finding,rombach2014core} and multi-layered networks \cite{gallagher2021clarified}.
In this present effort, we will not be measuring coreness of nodes or quantifying core-periphery structure in given networks.
Rather, we will generate networks at a desired level and type of core-periphery structure with an elementary random graph model and tie these networks to performance in our collective innovation task, as discussed in Sec.~\ref{subsection:preliminaries.SBMsCP}.

% \BLUE{This early paper anticipates later work in network statistics which would ground block modeling and core-periphery structure in a rigorous probabilistic framework} \RED{CITE}.

Ongoing work in the field of collective problem-solving has worked to identify network structures which optimize for the trade-off between the exploration of new solutions and the exploitation of current solutions \cite{centola2022network}. Emerging from these studies is an understanding on the role that informational efficiency in the form of simple path lengths plays in solving both simple and complex problems; with a core trade-off being that for simple problems, efficient or highly centralized graphs are desirable, while for more complex problems, inefficient and less clustered graphs are more desirable \cite{lazer2007network,cantor2021social}. Understood in these dimensions, networks which are efficient are said to be fast, with the ability to drive rapid consensus and "exploit" the information available to them while networks which are inefficient are said to be slower and more "transiently diverse", with the ability to engage in parallel exploration of different opportunities \cite{centola2022network,smaldino2022maintaining}. Thus, nearly networks in these tasks must work to strike the appropriate balance between optimal levels of exploration of the solution space and exploitation, \textit{apropos} their task \cite{march1991exploration}.

% \RED{Two Truths paper}
Many of the formal elements of our network analysis, including the generation of random networks from stochastic block models, the parameter distinction between core-periphery and affinity networks, and our spectral graph embedding analysis, draw from the work in \cite{priebe2019two}.
The authors compare core-periphery and affinity network structures from a spectral clustering perspective,
in order to explain the discrepancy between the communities assigned by ASE and LSE-based community detection methods.
They describe a ``Two Truths'' phenomenon, in which ASE preferentially reveals and clusters core-periphery communities, while LSE does the same for affinity communities.
In particularly striking results, they applied ASE and LSE clustering to a diffusion MRI connectome dataset, demonstrating that ASE captured grey/white matter core-periphery structure, while LSE captured left/right hemisphere affinity structure.
To our knowledge, the implications of this Two Truths hypothesis for spectral graph resampling \cite{levin2019bootstrapping} have not been explored.
There is also little work using spectral graph embeddings to analyze agent-based models, although they are commonplace in social network analysis. 

%%%%%%%%%%%%%%%%%%%%%%%%%%%%%%%%%%%%%%%%%%%%%%%%%%%%%%%%%%%%%%%%%%%%%%%%

\section{Preliminaries}
\label{section:preliminaries}

In this section, we introduce the formal background and technical methods of our investigation.
We use the terms `network' and `graph' interchangeably, in both cases always referring to undirected, unweighted graphs without loops.
These are pairs $\mathcal{G} = (\mathcal{V},\mathcal{E})$, where $\mathcal{V}$ is the set of vertices or nodes, and $\mathcal{E}$ is the set of edges.
We will always identify vertices with their indices, i.e. if $|\mathcal{V}| = N$, $\mathcal{V} = [N]$ where $[N] = \{ 1, ..., N\}$. An edge is a couple $\{i, j\} \in \mathcal{E}$ for some $i \neq j \in \mathcal{V}$.

\subsection{Stochastic Block Models of Affinity and Core-Periphery Networks}
\label{subsection:preliminaries.SBMsCP}

Similar to the influential work of Borgatti and Everett \cite{borgatti2000models}, we understand core-periphery structure as a network approximating a two-block model, with a heavily connected core and a less connected periphery.
We have opted to follow the loosely formalized probabilistic interpretation of core-periphery structure from Priebe \textit{et al.} \cite{priebe2019two}, in which they present core-periphery structure in contradistinction to affinity community structure under certain parameter regimes for 2-block stochastic block models (SBMs).
We expand this framework to include networks exhibiting the centralized core-periphery structure typical in the social networks literature, first formalized by \cite{borgatti2000models}.
As we will see, in this setting, the notions of core-periphery from \cite{borgatti2000models} and \cite{priebe2019two} are in some sense mutually exclusive, although they both assume the existence of a `rich club' core.

Stochastic block models (SBMs) are a popular model of random graphs (graph-valued random variables) with community structure, and are frequently used as the theoretical graph model underlying community detection algorithms within the network statistics literature, e.g. \cite{sussman2012consistent,lyzinski2016community}.
They are defined as follows.
Consider a collection of $N$ nodes divided into $K\geq 2$ blocks (`communities'), where each block $k$ is of size $n_k$, so that $\sum n_k = N$.
Let $\bm{\pi} \in [K]^N$ be the block membership vector.
The block matrix $\bm{B}$ is a $K \times K$ symmetric matrix of probabilities that parametrizes the SBM (along with $\bm{\pi}$), where $b_{k_1 k_2}$ provides the probability of an edge between a node in community $k_1$ and one in $k_2$.
This in turn forms an $N \times N$ edge probability matrix $\bm P$, composed of blocks $\bm P_{k_1 k_2} = b_{k_1 k_2} \mathds{1}_{k_i \times k_j}$.
The SBM is the random graph in which every edge $(i,j)$ is sampled independently, with probability $p_{ij}$.
We denote an SBM as a distribution on adjacency matrices: $A \sim \SBM(\bm B,\pi)$
Such a $P$-parametrized random graph model, more generally, is typically referred to as an inhomogenous Erdos-Renyi graph, denoted distributionally as $\bm A \sim \ER(\bm P)$.

Priebe \textit{et al.} \cite{priebe2019two} suggest a model for core-periphery networks as a 2-block SBM.
A two-block SBM is parametrized by the block matrix:
\begin{equation}
\label{eq:twoBlockMatrix}
\bm B =
\begin{bmatrix}
a & b \\ b & c
\end{bmatrix}
\end{equation}
They distinguish two generic cases of such SBMs. When $a,b \gg c$, the authors identify the SBM as exhibiting \textbf{affinity} structure, in which realizations of the network can usually be partitioned into two communities exhibiting dense intra-communal connectivity, with sparser connections between the two. In this case, neither community can be said to necessarily significantly dominate the other, especially if $a \approx c$ and $n_1 \approx n_2$. By contrast, the authors identified the case in which $a \gg b,c$ as exhibiting \textbf{core-periphery} structure. In this case, the first community exhibits dense intra-communal connectivity. By contrast, it is only weakly connected to the second community of peripheral nodes, which are sparsely connected among themselves. In this model of core-periphery structure, peripheral nodes, individually and communally, are isolated from the core as much as each other.

This understanding of core-periphery structure from \cite{priebe2019two} is similar to that of \cite{borgatti2000models} in that it understands core-periphery networks through a two-block model, composed of a dense core and sparse periphery.
However, there is a major conceptual distinction in the two works that is deeper than the difference in mathematical framing.
The approach in \cite{borgatti2000models} identifies an ideal core-periphery network as one in which each peripheral node is connected to every core node.
If we translated this into the more general probabilistic model (\ref{eq:twoBlockMatrix}), this would be the third case in which $a,b \gg c$, which was not considered in \cite{priebe2019two}.
In the case when $a,b \gg c$, peripheral nodes will be heavily connected to the core, while isolated from each other.
Within the context of our model, this would make it far more likely that peripheral nodes interact with and are influenced by the network core.
By contrast, when $a \gg b,c$, the peripheral nodes will be relatively isolated from both communities as such, with only a handful of neighbors.

Thus, we distinguish these two forms of core-periphery structure, referring to that from \cite{priebe2019two} as \textbf{decentralized core-periphery (DCP)} structure, and that from \cite{borgatti2000models} as \textbf{centralized core-periphery (CCP)} structure.
We refer to `centralization' to imply a notion of control on the part of the core, which is less evident in the core-periphery structure described by \cite{priebe2019two} than that in \cite{borgatti2000models}.
% \footnote{Notably, in \cite{priebe2019two}, core/periphery was mapped onto white/grey matter in the brain as a test case. It is far from clear that white matter regions can be understood as controling the functionining of grey matter, as grey matter regions are identified with most of the functionality we intuitively identify with agency in humans.}.
To summarize, there are three qualitatively distinct parameter regimes for (\ref{eq:twoBlockMatrix}) that we will be examining in this work:

\begin{figure}
\vspace{-1.5cm}
\includegraphics[scale=0.3]{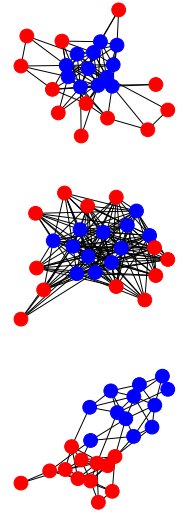}
\vspace{-1.5cm}
\caption{ \small \textbf{2-Block Stochastic Block Model Networks.} We generated three networks from model (\ref{eq:twoBlockMatrix}), with parameter sets corresponding to the three structural classes that we will be studying. All networks have $N=24$ nodes, divided evenly between the blocks ($n_1=n_2=12$). Blue nodes belong to the first block, while red nodes belong to the second. (A) A decentralized core-periphery network, generated with parameters $(a,b,c) = (0.8, 0.15, 0.05)$. As can be seen, the peripheral nodes have fewer neighbors than the core nodes, which include both core nodes and other peripheral nodes. (B) A centralized core-periphery network, generated with parameters $(a,b,c) = (0.8, 0.8, 0.05)$. (C) An affinity network, generated with parameters $(a,b,c) = (0.5, 0.5, 0.05)$.}
\label{fig:sample_networks}
\end{figure}
% p1s = [0.8, 0.8, 0.5]
% p2s = [0.05, 0.05, 0.5]
% qs = [0.15, 0.8, 0.05]

\begin{itemize}
\item \textbf{Affinity: $a,b \gg c$}. In affinity networks, there are two distinguishable block communities with more intra-block edges than inter-block edges. The level of intra-block connectivity is comparable between the two communities.
\item \textbf{Decentralized Core-Periphery (DCP): $a \gg b,c$}. There is a distinguishable core community, with dense intra-core connectivity.
The rest of the nodes are peripheral, with only a few connections to other nodes, core or peripheral.
\item \textbf{Centralized Core-Periphery (CCP): $a,b \gg c$}. There is a distinguishable core community, with dense intra-core connectivity \textit{and} dense connectivity to peripheral nodes. Peripheral nodes are heavily connected to the core, and only loosely connected to each other.
\end{itemize}

We hypothesize that in our ABM, DCP networks will have a performance advantage compared to affinity networks, as peripheral nodes in CCP networks are able to freely ``explore'' possible alternative combinations in the Potions Game while the core exploits whichever innovation pathway they are on. Additionally, few nodes in an affinity network will be sufficiently isolated to explore alternatives with high probability. We similarly predict that in CCP structures, peripheral nodes conform more to the behavior of the core and therefore limit diversity in the network, making the path-dependent nature of the task difficult to overcome, leading to an advantage of DCP networks over CCP networks.

\vspace{6pt}
\textbf{\hypertarget{hypothesis1}{Hypothesis 1}:} Networks exhibiting decentralized core-periphery structure will outperform those with either affinity or centralized core-periphery structure in collective innovation tasks.
\vspace{6pt}

For all networks, we will set the two blocks to the same size ($n = n_1 = n_2 = N/2$).
When comparing different parameter regimes for model (\ref{eq:twoBlockMatrix}), we want to maintain the same number of expected edges.
The relative density/sparsity of networks has been implicated \cite{lazer2007network,centola2022network} in collective problem-solving tasks, and we want to distinguish this effect from that of our competing core-periphery structures.
For a given 2-SBM of size $N=2n$, the expected number of edges is given by
\begin{equation}
\mathbb{E} |\mathcal{E}| = {n \choose 2} (a + b + c)
\end{equation}
Thus, to fix the total number of edges in the network, we need only fix the sum $a+b+c$.
Where possible, we compare networks generated by a linear family of SBMs:
\begin{equation}
    (a,b,c)(\theta) = (a_0,b_0,c_0) + \theta (\delta a, \delta b, \delta c), \; 0 \leq \theta \leq \theta_{\max}
\end{equation}
where we require $\delta a + \delta b + \delta c = 0$.
The only exception to this will be the family (\ref{eq:dcp_to_ccp_naive}), where we must allow for greater network density in order to make a DCP-CCP comparison without compromising on the density of the core, i.e. in order to raise $b$ without lowering $a$.

% Thus, for a given fixed density (average degree) $\rho = 2 |\mathcal{E}|/N$, we will compare parameter sets within the simplex given by $\mathbb{E} |\mathcal{E}| = N*\rho/2$.
% \todo[inline]{maybe add more with parameter sets, in terms of $\theta$-parametrization}

% It has the density function
% \begin{align}
% \nonumber P(A = M) &= \prod \left( p_{ij} m_{ij} + (1-p_{ij})(1 - m_{ij}) \right) \\
% &= \prod \left( b_{c_i c_j} m_{ij} + (1-b_{c_ic_j})(1 - m_{ij}) \right).
% \end{align}
% where $\bm{c}$ is the block membership vector.

\subsection{Simulation of Innovation Networks with Agent-Based Modeling}

We adapted an ABM of cumulative innovation from \cite{migliano2020hunter,cantor2021social}, derived from the potion-mixing game studied by Derex and Boyd in human players \cite{derex2016partial}. The only parameter varying between simulations is the structure of the agent network.

\begin{enumerate}
\item \textbf{Model initialization:} A network is created with its respective parameters. For each node of the graph, an agent is initialized with a score of zero and an inventory comprised of six items: three from an “A trajectory” and three from a “B trajectory.” Each item in the inventory is comprised of three parts: the name/level of the item (e.g., a1, a2, a3, b1, b2, b3) and a score which each initial item and items discovered thereafter carries for itself (with scores of 6, 8, and 10 for the three initial items in each trajectory).
\item \textbf{Dyad selection:} At each step, each agent chooses a partner they are connected to on the network with a random probability. As neighbors are simply chosen with a random probability, it is possible for a focal neighbor to select an individual which is already interacting with them (e.g., if a network is initialized with just two agents, the two agents will simply select each other).
\item \textbf{Item selection:} In the model, new items are formed by triad combinations of old items. As triad combinations are made between dyads of agents, the focal agent randomly selects whether it will be trading either one item or two items with their partner. The focal agent and its partner then cycle through their respective inventories, assigning probabilities to each item in the array. This is obtained by summing the innovation scores of each item and dividing individual scores by each sum (e.g., the initial inventory innovation scores of 6, 8, 10, 6, 8, 10 will yield respective probabilities of .125, .167, .208, .125, .167, .208).
\item \textbf{Item combination:} Agents and their partners then select the number of items previously assigned to them in the last step, based on items' calculated probabilities and without replacement, and combine them. The combination is saved as a list and compared to lists of valid combinations copied directly from Derex and Boyd \cite{derex2016partial} (SI Appendix, section 1). If an invalid combination is made, nothing happens. If a valid combination is made, then the dyad adds a new innovation (with its own respective score) to their inventories.
\item \textbf{Innovation diffusion:} If a new innovation is added to the agents' inventories, both agents then check the inventories of all of their partners and spread it to neighbors which do not already possess it.
\item \textbf{Scoring:} Scores are then obtained for each agent based on the tier of discovery an agent has obtained: with the first tier yielding a score of 48, second tier 109, third tier 188, and the fourth tier (which requires a crossover from the A and B trajectory) being 358. The maximum score of an item in an agent's list is determined to be their overall score.
\item \textbf{End and Crossover:} The simulation ends either when it has reached 1000 steps or when the network has achieved a "crossover event," whereby the final inventions in both trajectories are themselves combined, indicating the network has discovered and united both paths of exploration.
\end{enumerate}

\subsection{Spectral Graph Embeddings, Resampling, and the Two Truths Hypothesis}
\label{subsection:preliminaries.ASELSE}
Central limit theorems exist suggesting that both ASE and LSE provide asymptotically consistent estimates of the latent positions of random dot product graphs (RDPGs), of which SBMs are a special case \cite{sussman2013consistent,tang2018limit}.
We want to ask whether these two forms of spectral embedding capture a meaningful distinction between core-periphery and affinity networks, especially with respect to their performance in our innovation task.

Priebe and \textit{et al.}
% , who have extensively investigated the statistical properties of ASE and LSE, including their consistency in estimating block membership in SBMs,
put forward a two-truths hypothesis comparing the performance of ASE+GMM and LSE+GMM as community detection tools \cite{priebe2019two}.
They hypothesize that the latent positions estimated by ASE tend to emphasize core-periphery community structure,  grouping core and periphery nodes near each other in Euclidean space.
By contrast, the LSE seems to emphasize affinity communities.
In the context of their work, they were interested in community labels, with ASE and LSE used to discover different but equally true community partitions in connectome networks with known ground truth.
However, as mentioned above, asymptotic results suggest that the embedded positions themselves estimate latent positions for RDPGs.

Thus, given a network, we may use ASE and LSE to parameterize an RDPG model, which in turn will allow us to resample new networks with similar structure to the original.
From the Two Truths hypothesis, we expect the ASE-RDPG model to better capture DCP structure in the original network, while LSE-RDPG will reproduce affinity structure.

We start by defining the ASE and LSE.
Let a graph $\mathcal{G} = (\mathcal{V},\mathcal{E})$ be given, with $N=|\mathcal{V}|$.
The adjacency $\bm A$ matrix is an $N \times N$ symmetric matrix with Boolean values, where $a_{ij} = 1$ iff $\{i,j\} \in \mathcal{E}$, with zeros along the diagonal.
The normalized Laplacian matrix is given as $\bm L = \bm D^{-1/2} \bm A \bm D^{-1/2}$, where $\bm D$ is the degree matrix, i.e. diagonal where $d_{ii} = |\{ e \in \mathcal{E} \, | \, i \in e \}|$.
Both $\bm A$ and $\bm L$ are positive semidefinite, and have exclusively nonnegative eigenvalues.
For either matrix $\bm M$, we will have the eigen decomposition:
\begin{equation}
\bm M = \bm U \bm S \bm U^T
\end{equation}
where $\bm S$ is the diagonal matrix of the (ordered) eigenvalues of $\bm M$, and the columns of $\bm U$ are the corresponding eigenvectors.
For a given embedding dimension $d < \text{rank}(\bm M)$, we may instead compute the rank $k$ approximation:
\begin{equation}
\bm M \approx \bm U_d \bm S_d \bm U_d^T
\end{equation}
where $\bm S_d$ is the $d \times d$ diagonal matrix of the largest $d$ eigenvalues of $\bm M$, and $\bm U_d$ is the $N \times d$ matrix of corresponding orthonormal eigenvectors.
There are reasons both methodological and computational for using a low-rank decomposition, which we do not discuss here \cite{udell2019big}.
% Statistically, we prefer a model much less complex than our data.
% Practically, we observe that large data matrices, network or feature data, often demonstrate only a few significant eigenvalues before a noticeable ``elbow,'' i.e. a large gap in the eigenvalue plot between the largest eigenvalues and the rest, which are comparatively negligible in magnitude \RED{CITE, maybe "why matrices are low-rank"}.
% A common interpretation is that higher rank information past the elbow is noise.
% Computationally, low-rank decompositions are far more tractable for massive datasets, and there are diminishing returns for any practical appication.
These eigenvectors span a $d$-dimensional space into which we will embed our nodes.
To choose the embedding dimension $d$, we use the standard profile liklihood approach from \cite{zhu2006automatic}.
We then form the matrix $\bm \hat{X} = \bm U_d \sqrt{\bm S_d}$, with rows $\bm \hat X_i$ interpreted as the embedded node positions in $\mathbb{R}^d$.
This full embedding process is implemented in the open source \texttt{graspologic} package \cite{chung2019graspy}.

Both the ASE and LSE can be used to generate new networks of similar structure by treating the embedded node positions as latent positions within a random dot product graph (RDPG) model.
The RDPG model generalizes SBMs.
A $d$-dimensional RDPG model is parametrized by the $N \times d$ matrix of latent positions $\bm X$, with row $X_i$ corresponding to the latent position of node $i$.
The RDPG is an inhomogenous Erdos-Renyi graph with edge probability matrix $\bm P = \bm X \bm X^T$, i.e. the probability of edge $\{i,j\}$ is given by $p_{ij} = X_i \cdot X_j$.

With this in mind, we present ASE-RDPG and LSE-RDPG resampling distributions.
These are the mappings from a given network $\mathcal{G}$ to the RDPG taking $\hat{X}$ as its latent positions, i.e. $\mathcal{G} \to \ER(\hat X \hat X^T)$, with $\hat X$ formed as above via ASE or LSE.
However, given the relatively small size of our networks, we found it necessary to adjust our latent positions in order to generate fully connected networks.
Suppose we are embedding a network with $M$ edges.
For a given edge probability matrix $\bm P = \bm \hat X \bm \hat X^T$, derived from either ASE or LSE, we may compute the expected number of edges as $\mathbb{E} e(P) = \sum_{j>i} p_{ij}$.
We then computed the adjusted edge probability matrix $\tilde P = (M/\mathbb{E} e(\bm P)) \, \bm P$, and used $\bm \tilde P$ to generate new networks in place of $\bm P$.
Thus, using $\bm{X}_{\bm{A}}$ and $\bm{X}_{\bm{L}}$ to denote ASE and LSE spectral embeddings, we define our adjusted ASE-RDPG and LSE-RDPG distributions:

\begin{align}
\ASERDPG (\mathcal{G}) &= \ER ( r_{\bm{A}} \bm X_{\bm{A}}  \bm {X_{\bm{A}}}^\top ) \\
\LSERDPG (\mathcal{G}) &= \ER ( r_{\bm{L}} \bm X_{\bm{L}}  \bm {X_{\bm{L}}}^\top ) \\
\nonumber \text{ where } r_{\cdot} &= \frac{|\mathcal{E}|}{\mathbb{E} e(\bm X_{\cdot} \bm X_{\cdot}^T) }
\end{align}

When generating new networks, we also discarded any that were not connected.

In \hyperlink{hypothesis1}{Hypothesis 1} of Sec.~\ref{subsection:preliminaries.SBMsCP}, we predicted that DCP networks will outperform affinity (and CCP) networks in innovation tasks.
Building on the Two Truths hypothesis from \cite{priebe2019two}, we hold that ASE-RDPG resampling will generate new networks with similar DCP structure to the original network, and likewise LSE-RDPG will recreate affinity structure.
Putting the positive claims of these two together, we offer the following hypothesis:

\vspace{6pt}
\textbf{\hypertarget{hypothesis2}{Hypothesis 2}:} Given a base network, ASE-RDPG resampling will recreate (decentralized) core-periphery structure in new networks, which will in turn better approximate the performance of the original SBM-generated network in our task, relative to LSE-RDPG resampling.
\vspace{6pt}

In order to determine which spectral resampling method better approximates the ABM performance of the base networks, we will compare distributions of discovery times from the original and resampled networks, using the Earth Mover's Distance (EMD).
For discrete, real-valued distributions $\hat{f}(x_i)$ and $\hat{g}(y_j)$, the EMD is defined as:
\begin{equation}
d_{\text{EMD}}(\hat{f}, \hat{g}) = \inf_{\gamma \in \Gamma(\hat{f},\hat{g})} \sum_{i,j} \gamma(x_i, y_j) \; |x_i-y_j|
\end{equation}
where $\Gamma(\hat{f},\hat{g})$ is the collection of all product distributions $\gamma(x_i,y_j)$ with $\hat{f}$ and $\hat{g}$ as marginal distributions.

% Suppose we have network $\mathcal{G}$ generated by $A \sim \SBM(\bm{B}(a,b,c))$ for some parameter set $(a,b,c)$.
% We simulate our ABM on this network $m$ times, providing us with an empirical distribution of crossover times $\hat{f}_{\mathcal{G}} (x)$: 
% \begin{equation}
%     \hat{f}_{\mathcal{G}} (x) = \sum_{i=1}^m \delta(x - x_i)
% \end{equation}
% Then, using either ASE-RDPG or LSE-RDPG, we resample $m$ new networks $\bm{\hat{\mathcal{G}}} = (\hat{\mathcal{G}}_j)$.
% %  $\bm { \mathcal{G}} = (\hat \mathcal{G}_j)$ from $\mathcal{G}$.
% We simulate each network on the ABM simulation once, giving us crossover times $(y_j)_j$ with empirical distribution:
% \begin{equation}
%     \hat{g}_{\bm{\hat{\mathcal{G}}}}(y) = \sum_{j=1}^m \delta(y - y_j).
% \end{equation}
% We exclude the subscripts on $\hat{f}$ and $\hat{g}$. The 1-Wasserstein distance, also known as the Earth Mover's distance, is defined for discrete real-valued distributions as:
% \begin{equation}
% d_{W^1}(\hat{f}, \hat{g}) = \inf_{\gamma \in \Gamma(\hat{f},\hat{g})} \sum_{i,j} \gamma(x_i, y_j) |x_i-y_j|
% \end{equation}
% where $\Gamma(\hat{f},\hat{g})$ is the collection of all product distributions with $\hat{f}$ and $\hat{g}$ as marginals in the first and second argument, respectively.

\section{Results}

\subsection{Decentralized core-periphery structure accelerates innovation crossover relative to affinity structure.}
\label{subsection:results.dcp_to_affinity}

We first wanted to examine whether decentralized core-periphery structure promotes innovation crossover.
In our first experiment, we generated networks along a parameter edge-level curve:
\begin{equation}
\tag{M1}
\label{eq:dcp_to_affinity}
(a,b,c)(\theta) = (a_0,b_0,c_0) + \theta (-1,0,1), \; 0 \leq \theta \leq \theta_{\max}
\end{equation}
We will choose $(a_0,b_0,c_0)$ as a non-edge case in the decentralized core-periphery region.
As $\theta$ increases, the discrepancy between intra-block connectivities decreases, until they assume an equitable affinity structure.
In our first experiment, we set $a_0=0.75, c_0=0.05$, and consider four values for inter-block probability $b = b_0 = 0.01, 0.05, 0.1,$ or $0.15$.
For each parameter regime, we generated 500 networks from the SBM, and ran the simulation on each network until the final innovation was discovered, recording the number of time steps needed for each.

\begin{figure}
% \centering
\vspace{-1.5cm}
% \includegraphics[scale=0.4]{figure_2SBM_CP_batch_3.eps}
% \hspace*{-.5cm}\includegraphics[scale=0.45]{figure_experiment_1_adjust_b_barplots.eps}
\hspace*{-.5cm}\includegraphics[scale=0.6]{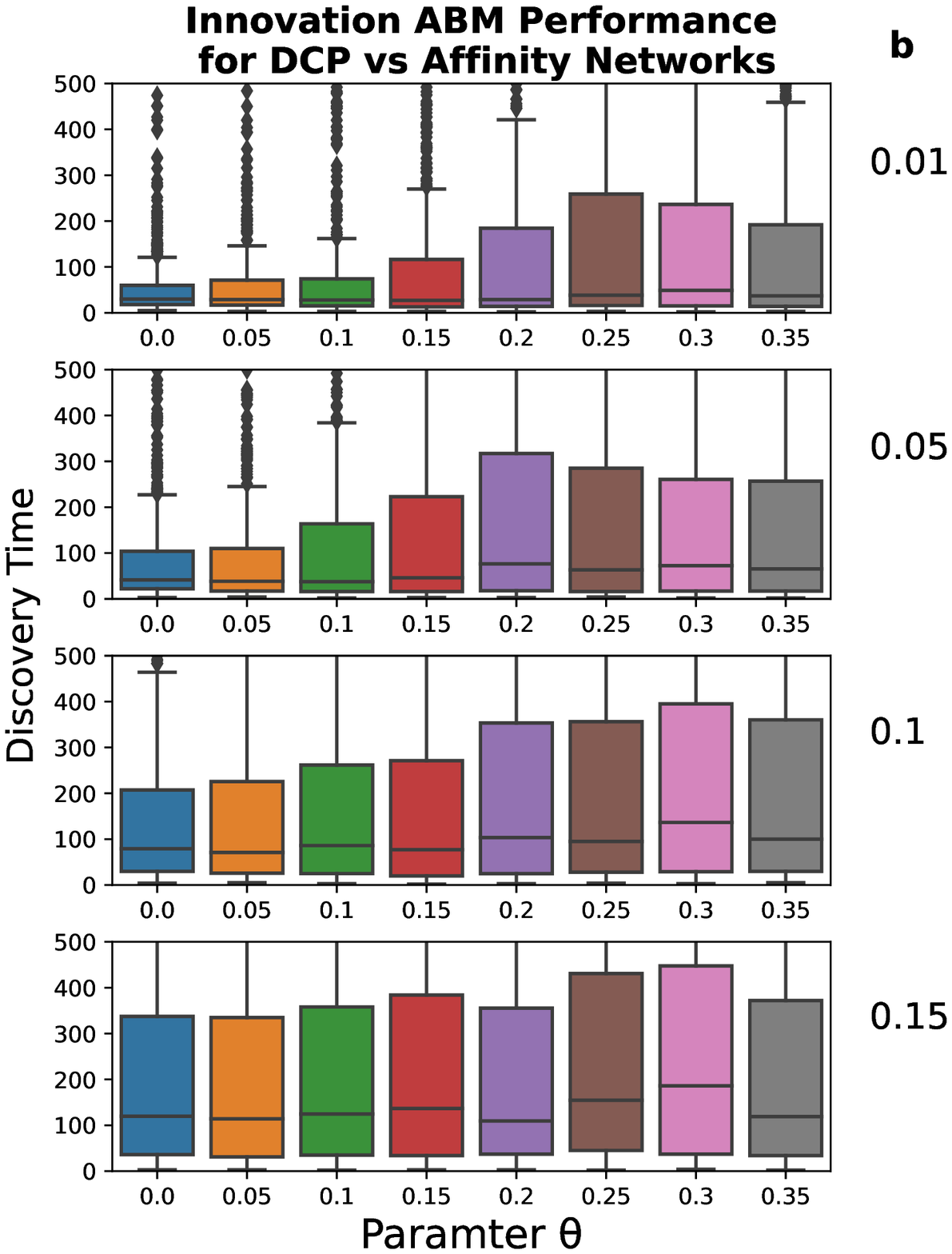}
\includegraphics[scale=0.55]{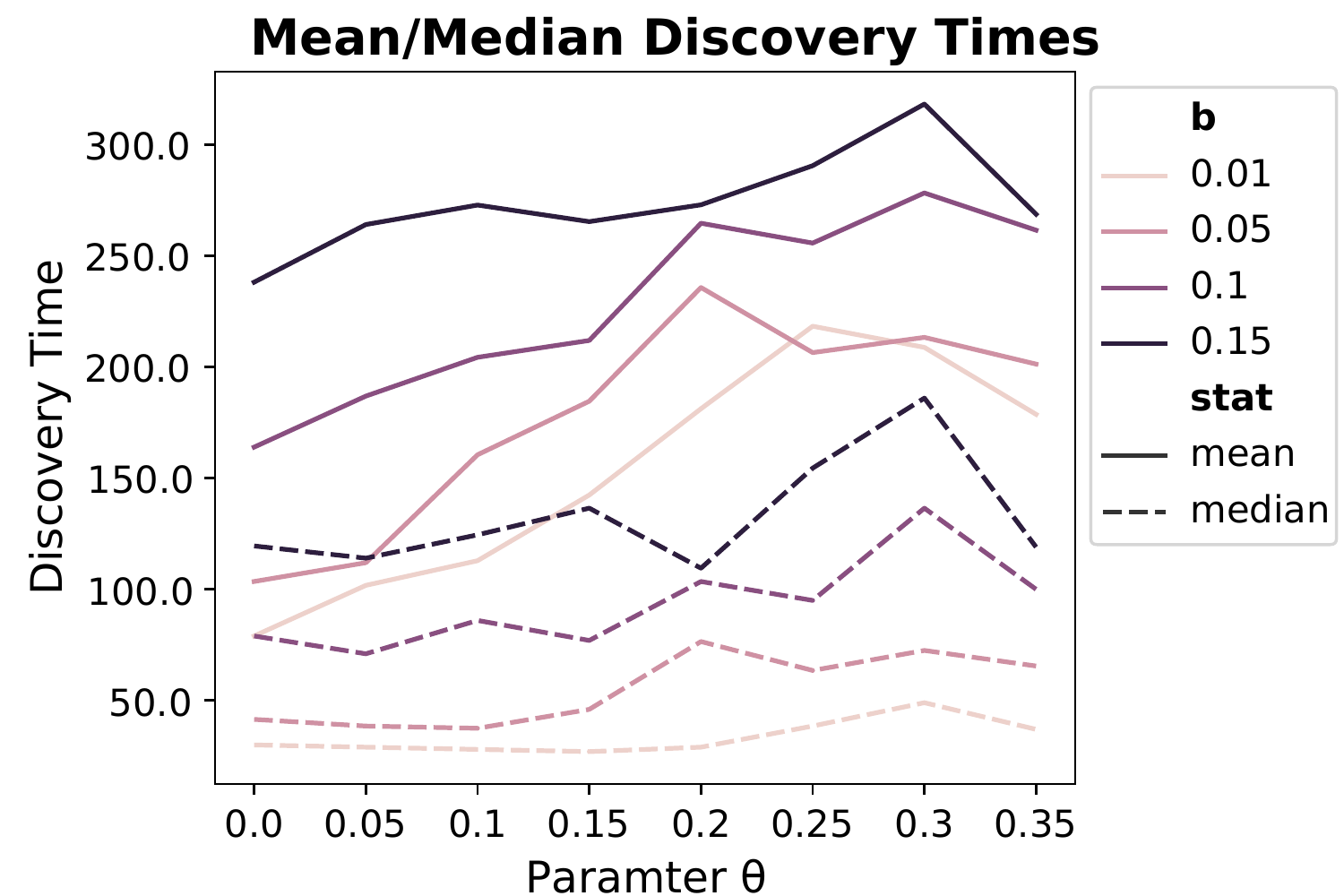}

\caption{(TOP) The distributions of crossover discovery times (in discrete time steps) for networks generated from the family of SBMs (\ref{eq:dcp_to_affinity}), parameterized by $\theta$.
For each row, $a_0=0.75,c_0=0.05,$ and core-periphery connectivity $b$ is held constant according to the value to the right of the plot.
Networks exhibit DCP structure for small $\theta$, and assume an affinity structure as $\theta$ grows.
(BOTTOM) We separately visualize the means and medians of the distributions from the boxplots.}
\label{fig:dcp_to_affinity}
\end{figure}

We present the results of this simulation experiment in Fig.~\ref{fig:dcp_to_affinity}.
We show the distribution of discovery times as we increase $\theta$, transitioning from decenralized core-periphery to affinity structure, for each choice of $b=b_0$.
Consider first the case $b_0=c_0 = 0.05$, where $(a,b,c,)(0) = (0.75, 0.05, 0.05)$ is a relatively balanced decentralized core-periphery SBM, with equal core-periphery and intra-periphery edge probabilities. 
As can be seen, the mean crossover time is lowest for the decentralized core-periphery networks generated at $\theta = 0$, supporting Hypothesis 1.
As $\theta$ increases, the mean crossover time likewise increases.
Interestingly, the median is more stable across the parameter curve.
The distribution of crossover times is non-Gaussian, and has a long right tail.
The tail `shrinks' as $\theta \to 0^+$, suggesting that decentralized core-periphery structure may accelerate innovation in expectation by reducing the likelihood of stagnation.

Now, consider this relationship between $\theta$ (DCP-to-affinity) and mean crossover time in light of the progression of distributions for all four values of $b$ in Fig.~\ref{fig:dcp_to_affinity}.
We see that this relationship is dependent upon the low value we chose for the inter-block parameter $b$.
The performance of core-periphery networks \textit{increases} if we decrease $b$ to $0.01$, further isolating the perphery from the core, while discovery times degrade rather quickly if we double or triple $b$.
This suggests that the accelerated innovation we observe in simulation for core-periphery social networks is dependent upon the network inefficiency that separates peripheral nodes from most of the core.

\subsection{Centralized core-periphery structure does not accelerate innovation crossover, relative to affinity structure.}
\label{subsection:results.ccp_to_affinity}
% \todo[inline]{All text in green refers to another experiment I ran, but will likely cut for time/space.}
The results presented in Fig.~\ref{fig:dcp_to_affinity} suggest that the advantage in discovery times observed in simulation for our decentralized core-periphery innovation networks is dependent on their decentralization, i.e. the low inter-block connectivity parametrized in our SBM by $b$.
This would cast doubt on any advantage in innovation speed for centralized core-periphery networks, in the sense of \cite{borgatti2000models}.

We considered this explicitly with two $\theta$-parametrized families of SBMs.
We thus turn to examining the performance of SBMs exhibiting CCP structure in our simulation.
Similar to before, we consider the transition from CCP to affinity structure in our SBM:
\begin{equation}
    \tag{M2}
    \label{eq:ccp_to_affinity}
    (a,b,c) = (a_0, b_0, c_0) + \theta \left(0,-1,1 \right)
\end{equation}
% {
% \color{green}
% We also considered a model family where all community structure degrades, and the model approximates a homogenous Erdos-Renyi random graph as we increase $\theta$:
% \begin{equation}
%   \tag{M3}
%   \label{eq:ccp_to_random}
%   (a,b,c) = (a_0, a_0, c_0) + \theta \left( -\frac{1}{2},-\frac{1}{2},1 \right)
% \end{equation}
% This latter model allows us to compare CCP networks to maximally random networks of the same (expected) edge count.
% }

\begin{figure}
% \centering
% \includegraphics[scale=0.4]{figure_2SBM_CP_batch_3.eps}
% \hspace*{-.5cm}\includegraphics[scale=0.55]{figure_CCP_to_Affinity.pdf}
\hspace*{-.5cm}\includegraphics[scale=0.55]{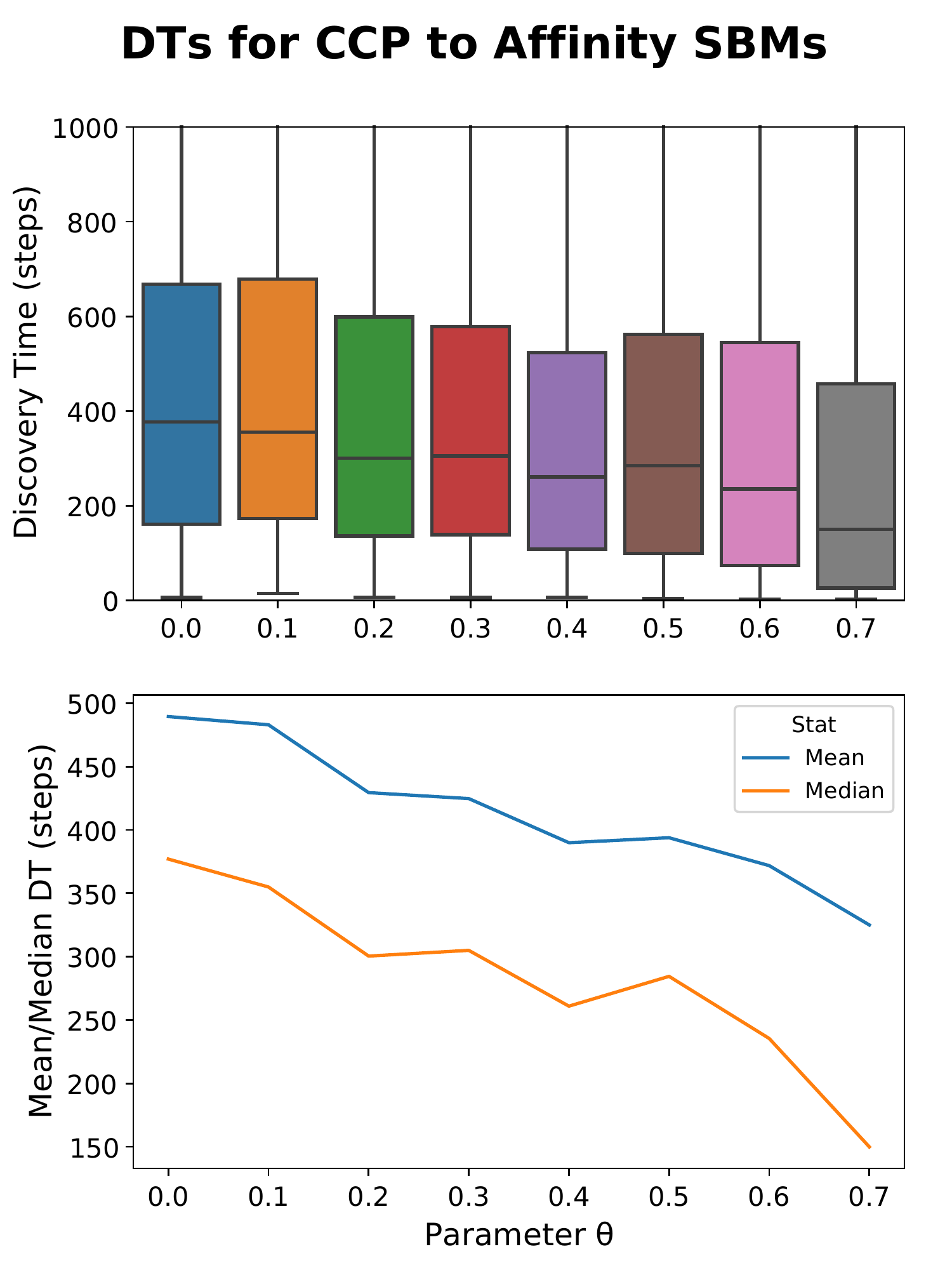}
\caption{The distributions of discovery times for networks generated from the family (M2), with $(a_0,b_0,c_0)=(0.75,0.75,0.05)$. Networks exhibit CCP structure for $\theta=0$, and assume an affinity structure as $\theta$ grows. As can be seen, mean and median discovery time decrease over the transition from CCP SBMs to affinity SBMs.}
\label{fig:CCP_to_affinity}
\end{figure}

We present the results of our simulations for (\ref{eq:ccp_to_affinity}) in Fig.~\ref{fig:CCP_to_affinity}. 
Networks exhibiting centralized core-periphery structure seem to realize no advantage in innovation tasks, relative to affinity networks composed two internally dense, loosely connected communities.
To the contrary, the affinity networks actually outperform CCP networks, in terms of lower mean and median discovery times.
These results indicate that from the perspective of our collective problem-solving, centralized and decentralized core-periphery networks are more different from each other than either is from a balanced affinity network, in terms of innovation discovery speed. 

\begin{figure}
% \centering
% \includegraphics[scale=0.4]{figure_2SBM_CP_batch_3.eps}
% \hspace*{-.5cm}\includegraphics[scale=0.55]{figure_compareCP_combined_plot.pdf}
\hspace*{-.5cm}\includegraphics[scale=0.65]{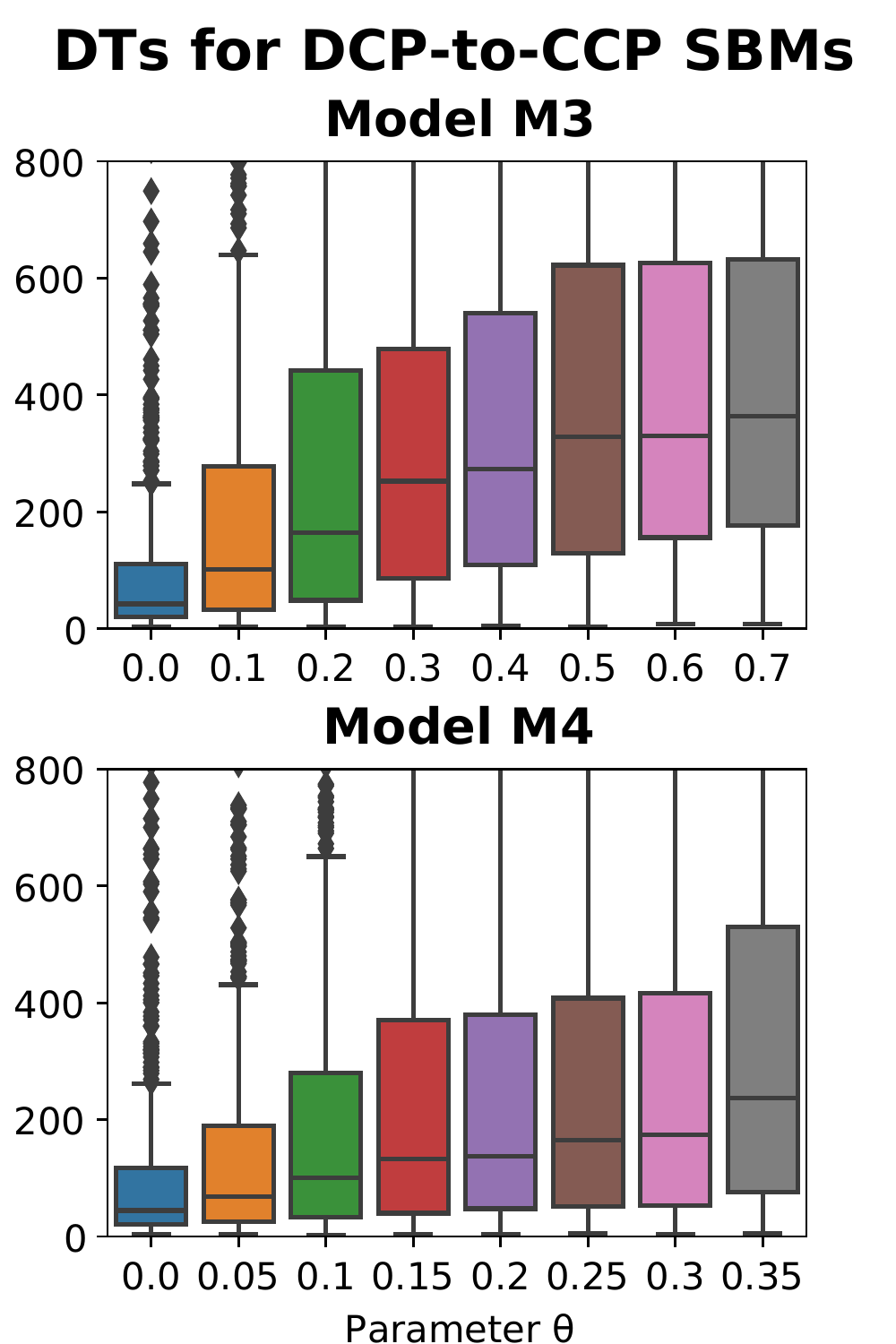}
\caption{The distribution of discovery times for networks generated from (\ref{eq:dcp_to_ccp_naive}) and (\ref{eq:dcp_to_ccp_balanced}), with $(a_0,b_0,c_0)=(0.75,0.05 , 0.05)$. Networks transition from DCP to CCP structure as $\theta$ increases. As can be seen, discovery time increases in both models as the core-periphery structure becomes centralized.}
\label{fig:compareCP_combined}
\end{figure}

\begin{figure}
% \includegraphics[scale=0.4]{figure_2SBM_CP_batch_3.eps}
% \hspace*{-.25cm}\includegraphics[scale=0.6]{figure_ASELSE.pdf}
\hspace*{-.25cm}\includegraphics[scale=0.6]{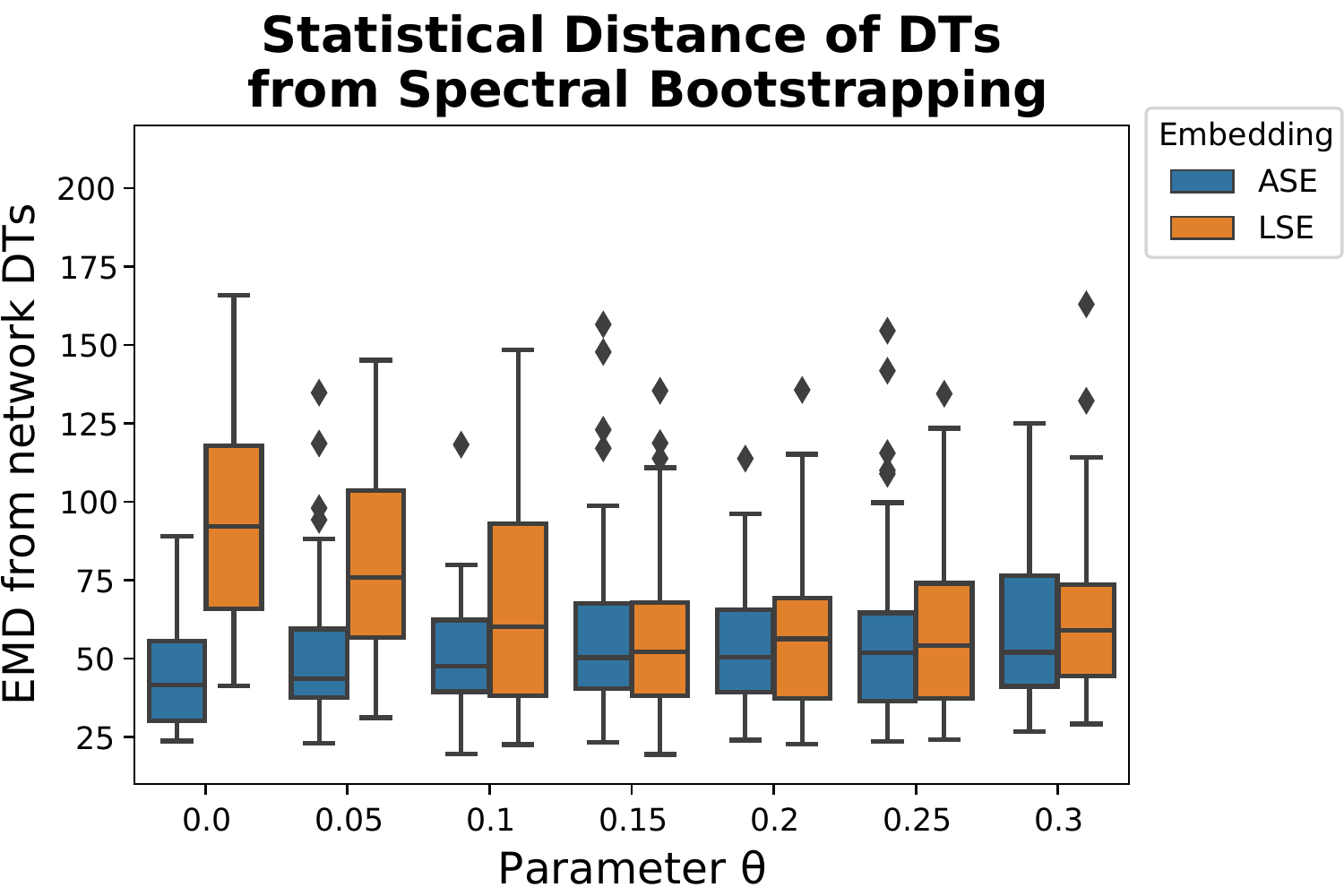}
\caption{Statistical distances between the discovery time distributions of model (\ref{eq:dcp_to_affinity}) networks (the original) and corresponding ASE/LSE-RDPG resampled networks. For $\theta=0$ (DCP networks), we see that the DT distribution of ASE-RDPG networks more closely matches that of the original network, in line with the Two Truths hypothesis from \cite{priebe2019two}.}
\label{fig:experimentSpectraEmbedding}
\end{figure}

% {
% \color{green}Next, we examine model (\ref{eq:ccp_to_random} in Fig.~(\RED{SUPP}) in the Supplemental Material. We set $a_0=0.75$ and $c_0=0.05$, and allowed $0 \leq \theta \leq 0.45$.
% As $\theta$ increases, the SBM approaches a Erdos-Renyi graph and loses its centralized core-periphery structure.
% We found no clear benefit for the centralized core-periphery structure upon mean or median discovery time (Fig~\RED{SUPP}).
% If anything, mean discovery time decreased as the network became more entropic, i.e. more uniformly random.
% This further reinforces the notion
% }

To conclude, our results in this section highlight longer discovery times exhibited by centralized core-periphery networks in comparison to affinity networks in the ABM. This suggests that the accelerated innovation observed in DCP networks requires both the decentralized ($a \gg b$) and core-periphery ($a \gg c$) elements of its structure. Moreover, our results highlight the qualitative distinction between the two types of core-periphery networks, which clearly has impact on the diffusion and diversity of information within the network.

\subsection{Decentralized core-periphery structure outperforms centralized core-periphery structure}

In Sec.~\ref{subsection:results.dcp_to_affinity} and Sec.~\ref{subsection:results.ccp_to_affinity}, we compared the innovation performance of DCP and CCP SBMs individually to affinity SBMs, finding DCP outperformed affinity networks, which are in turn outperformed CCP networks.
We round out these comparisons by running a direct comparison between the two types of core-periphery networks: decentralized and centralized.
A more difficult task is that of determining the appropriate categorical comparison between DCP and CCP networks: we expect an SBM with $a,b \gg c$ to have more edges and shorter path distances than one with $a \gg b, c$, assuming we hold $a$ and $c$ constant.
Thus, we will consider two families of SBMs, one preserving the expected number of edges at the expense of comparable core structure, and one allowing the CCP networks more edges in expectation in order to preserve comparable core density between CCP and DCP:
\begin{align}
\tag{M3}
\label{eq:dcp_to_ccp_naive}
(a,b,c) &= (a_0, b_0, c_0) + \theta \left(0,1,0 \right)\\
\tag{M4}
\label{eq:dcp_to_ccp_balanced}
(a,b,c) &= (a_0, b_0, c_0) + \theta \left(-1,1,0\right)
\end{align}
The family (\ref{eq:dcp_to_ccp_naive}) breaks with our convention up to this point, increasing the expected number of edges linearly in $\theta$ in order to transition from DCP to CCP structure.
The family (\ref{eq:dcp_to_ccp_balanced}) preserves average density, but at the cost of a less dense core.
Fortunately this mattered relatively little for the qualitative results.

We present the results of simulating both families of SBMs in Fig.~\ref{fig:compareCP_combined}. As can be seen, regardless of which parameter curve is taken, transitioning from DCP to CCP structure increases mean discovery times.

\subsection{Performance of ASE/LSE-resampled networks implicates (decentralized) core-periphery structure.}
\label{subsection:results.ASELSE}

We have demonstrated that DCP networks outperform both affinity and CCP networks in our ABM, supporting \hyperlink{hypothesis1}{Hypothesis 1}.
Recall that the distinction between DCP and affinity network structure, as previously articulated  in \cite{priebe2019two}, was motivated by a Two Truths phenomenon in spectral clustering, whereby both ASE- and LSE-based community detection found meaningful but distinct pairs of ground-truth communities in connectome networks.
In what we have referred to as the Two Truths hypothesis, the authors suggested that ASE captures (decentralized) core-periphery structure, while LSE captures affinity structure.
Thus, we posed \hyperlink{hypothesis2}{Hypothesis 2}, in which we speculated that networks generated by ASE-RDPG resampling would better reproduce the original ABM performance of a network exhibiting DCP structure, while those from LSE-RDPG resampling would do so for affinity networks.

To test this hypothesis, we again consider networks generated from the SBM family (\ref{eq:dcp_to_affinity}). We used the base DCP paramter set $(a_0,b_0,c_0) = (0.75, 0.05, 0.15)$.\footnote{We initially attempted to use the same base DCP parameter set $(a_0,b_0,c_0) = (0.75, 0.05, 0.05)$. Interestingly, even with the adjustments described in Sec.~\ref{subsection:preliminaries.ASELSE} to scale the latent positions to the desired expected degree, we found that the ASE-RDPG rarely generated connected networks from a connected network.} For each $\theta$, we generated 50 networks. For each network, we ran 100 simulations of the ABM, forming an empirical distribution $\hat{f}$ of crossover times. We then generated 100 networks each via ASE-RDPG and LSE-RDPG resampling, simulating our ABM once per resampled network. This provided us with empirical distributions $\hat{g}_{\text{ASE}}$ and $\hat{g}_{\text{LSE}}$ of discovery times.
Note that the randomness of $\hat{f}$ is purely that of the simulator, while both resampled $\hat{g}_{\cdot}$ have randomness associated with both the simulator and the network resampling itself.
From all three distributions, we compute the statistical distance $d_{\text{EMD}}(\hat{f}, \hat{g}_{\cdot})$ for each embedding (Sec.~\ref{subsection:preliminaries.ASELSE}).
We collected these distances for all 50 base networks for each $\theta$.

Figure~\ref{fig:experimentSpectraEmbedding} shows the distributions of the statistical distance between $\hat{f}$ and $\hat{g}_{\cdot}$ for each type of spectral embedding, grouped by $\theta$.
We see that for $\theta=0$, i.e. the SBM with the strongest DCP structure, the average distance between DTs from the SBM-generated network and those of the ASE-RDPG resampled networks ($d_{\text{EMD}}(\hat{f} , \hat{g}_{\text{ASE}})$) was significantly less than the distance to the DT distribution of the LSE-RDPG networks ($d_{\text{EMD}}(\hat{f},\hat{g}_{\text{LSE}})$.
This suggests that, in line with the Two Truths hypothesis, ASE-RDPG is preferentially capturing and recreating the DCP structure of the original network, which in turn is responsible for more similar network performance in the ABM simulations.
This supports Hypothesis 2.

However, we also see that when $\theta=0.35$, i.e. $(a,b,c)=(0.45, 0.05, 0.45)$ and the SBM is generating affinity networks, ASE-RDPG and LSE-RDPG resampling recreate the DTs of the original network to a similar degree. That is to say, neither is preferentially producing networks closer in performance to the base SBM network.
If we may take the Two Truths hypothesis for granted, our results in this section might suggest that the performance of DCP networks is due to DCP structure, but the performance of affinity networks is less reflective of affinity community structure as such.
Alternatively, it is likely that the Two Truths phenomenon observed by \cite{priebe2019two} is a tendency, but the network features captured by each embedding are far more subtle than DCP versus affinity. In other words, the Two Truths hypothesis should be given less weight.

\section{Discussion}
Our investigation highlights the role that core-periphery structure plays in a network's ability to solve a collective problem-solving task.
Furthermore, it has provided an operational understanding of the topological distinction between two forms of core-periphery structure, that of decentralized core-periphery (DCP) structure and that of centralized core-periphery (CCP) structure.

% Using agent-based modeling, our investigation highlights the role that core-periphery structure plays in a network's ability to solve a collective problem-solving task and furthermore help to distinguish two forms of core-periphery structure, that of decentralized core-periphery (DCP) structure and that of centralized core-periphery (CCP) structure.

Utilizing an agent-based model in which populations must maintain a balance between exploring novel combinations and exploiting combinations they have, we showed that DCP networks, in which peripheral nodes are isolated from both a centralized core and one another, outperform affinity networks; that this same relationship is not found between affinity and CCP networks, in which peripheral nodes are isolated from one another but still maintain close links with the core; and that in a direct comparison, DCP networks outperform CCP networks.

In addition to this contribution to the literature on collective problem solving, we have investigated the implications of using different spectral graph embeddings for the resampling of networks in agent-based modeling.
This investigation has provided an operational and dynamic significance to a graph-theoretic phenomenon, in this case the Two Truths hypothesis for spectral clustering \cite{priebe2019two}.
By sampling generated networks and parameterizing based on their ASE, which is hypothesized to capture core-periphery structure, and LSE, which is hypothesized to capture affinity structure, and comparing them to performance in our original DCP networks, we found that our ASE generated networks performed most similarly to our DCP networks, while the LSE networks did not.

\clearpage

%%%%%%%%%%%%%%%%%%%%%%%%%%%%%%%%%%%%%%%%%%%%%%%%%%%%%%%%%%%%%%%%%%%%%%%%

%%% The acknowledgments section is defined using the "acks" environment
%%% (rather than an unnumbered section). The use of this environment 
%%% ensures the proper identification of the section in the article 
%%% metadata as well as the consistent spelling of the heading.

%%%%%%%%%%%%%%%%%%%%%%%%%%%%%%%%%%%%%%%%%%%%%%%%%%%%%%%%%%%%%%%%%%%%%%%%

%%% The next two lines define, first, the bibliography style to be 
%%% applied, and, second, the bibliography file to be used.

\bibliographystyle{ACM-Reference-Format} 
\bibliography{references}

%%% -*-BibTeX-*-
%%% Do NOT edit. File created by BibTeX with style
%%% ACM-Reference-Format-Journals [18-Jan-2012].

\begin{thebibliography}{32}

%%% ====================================================================
%%% NOTE TO THE USER: you can override these defaults by providing
%%% customized versions of any of these macros before the \bibliography
%%% command.  Each of them MUST provide its own final punctuation,
%%% except for \shownote{}, \showDOI{}, and \showURL{}.  The latter two
%%% do not use final punctuation, in order to avoid confusing it with
%%% the Web address.
%%%
%%% To suppress output of a particular field, define its macro to expand
%%% to an empty string, or better, \unskip, like this:
%%%
%%% \newcommand{\showDOI}[1]{\unskip}   % LaTeX syntax
%%%
%%% \def \showDOI #1{\unskip}           % plain TeX syntax
%%%
%%% ====================================================================

\ifx \showCODEN    \undefined \def \showCODEN     #1{\unskip}     \fi
\ifx \showDOI      \undefined \def \showDOI       #1{#1}\fi
\ifx \showISBNx    \undefined \def \showISBNx     #1{\unskip}     \fi
\ifx \showISBNxiii \undefined \def \showISBNxiii  #1{\unskip}     \fi
\ifx \showISSN     \undefined \def \showISSN      #1{\unskip}     \fi
\ifx \showLCCN     \undefined \def \showLCCN      #1{\unskip}     \fi
\ifx \shownote     \undefined \def \shownote      #1{#1}          \fi
\ifx \showarticletitle \undefined \def \showarticletitle #1{#1}   \fi
\ifx \showURL      \undefined \def \showURL       {\relax}        \fi
% The following commands are used for tagged output and should be
% invisible to TeX
\providecommand\bibfield[2]{#2}
\providecommand\bibinfo[2]{#2}
\providecommand\natexlab[1]{#1}
\providecommand\showeprint[2][]{arXiv:#2}

\bibitem[\protect\citeauthoryear{Bassett, Wymbs, Rombach, Porter, Mucha, and
  Grafton}{Bassett et~al\mbox{.}}{2013}]%
        {bassett2013task}
\bibfield{author}{\bibinfo{person}{Danielle~S Bassett},
  \bibinfo{person}{Nicholas~F Wymbs}, \bibinfo{person}{M~Puck Rombach},
  \bibinfo{person}{Mason~A Porter}, \bibinfo{person}{Peter~J Mucha}, {and}
  \bibinfo{person}{Scott~T Grafton}.} \bibinfo{year}{2013}\natexlab{}.
\newblock \showarticletitle{Task-based core-periphery organization of human
  brain dynamics}.
\newblock \bibinfo{journal}{\emph{PLoS Computational Biology}}
  \bibinfo{volume}{9}, \bibinfo{number}{9} (\bibinfo{year}{2013}),
  \bibinfo{pages}{e1003171}.
\newblock


\bibitem[\protect\citeauthoryear{Borgatti and Everett}{Borgatti and
  Everett}{2000}]%
        {borgatti2000models}
\bibfield{author}{\bibinfo{person}{Stephen~P Borgatti} {and}
  \bibinfo{person}{Martin~G Everett}.} \bibinfo{year}{2000}\natexlab{}.
\newblock \showarticletitle{Models of core/periphery structures}.
\newblock \bibinfo{journal}{\emph{Social Networks}} \bibinfo{volume}{21},
  \bibinfo{number}{4} (\bibinfo{year}{2000}), \bibinfo{pages}{375--395}.
\newblock


\bibitem[\protect\citeauthoryear{Cantor, Chimento, Smeele, He, Papageorgiou,
  Aplin, and Farine}{Cantor et~al\mbox{.}}{2021}]%
        {cantor2021social}
\bibfield{author}{\bibinfo{person}{Mauricio Cantor}, \bibinfo{person}{Michael
  Chimento}, \bibinfo{person}{Simeon~Q Smeele}, \bibinfo{person}{Peng He},
  \bibinfo{person}{Danai Papageorgiou}, \bibinfo{person}{Lucy~M Aplin}, {and}
  \bibinfo{person}{Damien~R Farine}.} \bibinfo{year}{2021}\natexlab{}.
\newblock \showarticletitle{Social network architecture and the tempo of
  cumulative cultural evolution}.
\newblock \bibinfo{journal}{\emph{Proceedings of the Royal Society B}}
  \bibinfo{volume}{288}, \bibinfo{number}{1946} (\bibinfo{year}{2021}),
  \bibinfo{pages}{20203107}.
\newblock


\bibitem[\protect\citeauthoryear{Cattani and Ferriani}{Cattani and
  Ferriani}{2008}]%
        {cattani2008core}
\bibfield{author}{\bibinfo{person}{Gino Cattani} {and} \bibinfo{person}{Simone
  Ferriani}.} \bibinfo{year}{2008}\natexlab{}.
\newblock \showarticletitle{A core/periphery perspective on individual creative
  performance: Social networks and cinematic achievements in the Hollywood film
  industry}.
\newblock \bibinfo{journal}{\emph{Organization Science}} \bibinfo{volume}{19},
  \bibinfo{number}{6} (\bibinfo{year}{2008}), \bibinfo{pages}{824--844}.
\newblock


\bibitem[\protect\citeauthoryear{Cattani, Ferriani, and Colucci}{Cattani
  et~al\mbox{.}}{2015}]%
        {cattani2015creativity}
\bibfield{author}{\bibinfo{person}{Gino Cattani}, \bibinfo{person}{Simone
  Ferriani}, {and} \bibinfo{person}{Mariachiara Colucci}.}
  \bibinfo{year}{2015}\natexlab{}.
\newblock \showarticletitle{Creativity in Social Networks}.
\newblock In \bibinfo{booktitle}{\emph{The Oxford handbook of creative
  industries}}. \bibinfo{publisher}{Oxford University Press Oxford},
  \bibinfo{pages}{75--95}.
\newblock


\bibitem[\protect\citeauthoryear{Centola}{Centola}{2021}]%
        {centola2021influencers}
\bibfield{author}{\bibinfo{person}{Damon Centola}.}
  \bibinfo{year}{2021}\natexlab{}.
\newblock \showarticletitle{Influencers, Backfire Effects, and the Power of the
  Periphery}.
\newblock \bibinfo{journal}{\emph{Personal Networks: Classic Readings and New
  Directions in Egocentric Analysis}}  \bibinfo{volume}{51}
  (\bibinfo{year}{2021}), \bibinfo{pages}{73}.
\newblock


\bibitem[\protect\citeauthoryear{Centola}{Centola}{2022}]%
        {centola2022network}
\bibfield{author}{\bibinfo{person}{Damon Centola}.}
  \bibinfo{year}{2022}\natexlab{}.
\newblock \showarticletitle{The network science of collective intelligence}.
\newblock \bibinfo{journal}{\emph{Trends in Cognitive Sciences}}
  (\bibinfo{year}{2022}).
\newblock


\bibitem[\protect\citeauthoryear{Chung, Pedigo, Bridgeford, Varjavand, Helm,
  and Vogelstein}{Chung et~al\mbox{.}}{2019}]%
        {chung2019graspy}
\bibfield{author}{\bibinfo{person}{Jaewon Chung}, \bibinfo{person}{Benjamin~D
  Pedigo}, \bibinfo{person}{Eric~W Bridgeford}, \bibinfo{person}{Bijan~K
  Varjavand}, \bibinfo{person}{Hayden~S Helm}, {and} \bibinfo{person}{Joshua~T
  Vogelstein}.} \bibinfo{year}{2019}\natexlab{}.
\newblock \showarticletitle{GraSPy: Graph Statistics in Python.}
\newblock \bibinfo{journal}{\emph{J. Mach. Learn. Res.}} \bibinfo{volume}{20},
  \bibinfo{number}{158} (\bibinfo{year}{2019}), \bibinfo{pages}{1--7}.
\newblock
\urldef\tempurl%
\url{https://github.com/microsoft/graspologic/}
\showURL{%
\tempurl}


\bibitem[\protect\citeauthoryear{Clune, Mouret, and Lipson}{Clune
  et~al\mbox{.}}{2013}]%
        {clune2013evolutionary}
\bibfield{author}{\bibinfo{person}{Jeff Clune}, \bibinfo{person}{Jean-Baptiste
  Mouret}, {and} \bibinfo{person}{Hod Lipson}.}
  \bibinfo{year}{2013}\natexlab{}.
\newblock \showarticletitle{The evolutionary origins of modularity}.
\newblock \bibinfo{journal}{\emph{Proceedings of the Royal Society b:
  Biological sciences}} \bibinfo{volume}{280}, \bibinfo{number}{1755}
  (\bibinfo{year}{2013}), \bibinfo{pages}{20122863}.
\newblock


\bibitem[\protect\citeauthoryear{Derex and Boyd}{Derex and Boyd}{2016}]%
        {derex2016partial}
\bibfield{author}{\bibinfo{person}{Maxime Derex} {and} \bibinfo{person}{Robert
  Boyd}.} \bibinfo{year}{2016}\natexlab{}.
\newblock \showarticletitle{Partial connectivity increases cultural
  accumulation within groups}.
\newblock \bibinfo{journal}{\emph{Proceedings of the National Academy of
  Sciences}} \bibinfo{volume}{113}, \bibinfo{number}{11}
  (\bibinfo{year}{2016}), \bibinfo{pages}{2982--2987}.
\newblock


\bibitem[\protect\citeauthoryear{Gallagher, Young, and Welles}{Gallagher
  et~al\mbox{.}}{2021}]%
        {gallagher2021clarified}
\bibfield{author}{\bibinfo{person}{Ryan~J Gallagher},
  \bibinfo{person}{Jean-Gabriel Young}, {and} \bibinfo{person}{Brooke~Foucault
  Welles}.} \bibinfo{year}{2021}\natexlab{}.
\newblock \showarticletitle{A clarified typology of core-periphery structure in
  networks}.
\newblock \bibinfo{journal}{\emph{Science advances}} \bibinfo{volume}{7},
  \bibinfo{number}{12} (\bibinfo{year}{2021}), \bibinfo{pages}{eabc9800}.
\newblock


\bibitem[\protect\citeauthoryear{Kojaku and Masuda}{Kojaku and Masuda}{2017}]%
        {kojaku2017finding}
\bibfield{author}{\bibinfo{person}{Sadamori Kojaku} {and}
  \bibinfo{person}{Naoki Masuda}.} \bibinfo{year}{2017}\natexlab{}.
\newblock \showarticletitle{Finding multiple core-periphery pairs in networks}.
\newblock \bibinfo{journal}{\emph{Physical Review E}} \bibinfo{volume}{96},
  \bibinfo{number}{5} (\bibinfo{year}{2017}), \bibinfo{pages}{052313}.
\newblock


\bibitem[\protect\citeauthoryear{Lazer and Friedman}{Lazer and
  Friedman}{2007}]%
        {lazer2007network}
\bibfield{author}{\bibinfo{person}{David Lazer} {and} \bibinfo{person}{Allan
  Friedman}.} \bibinfo{year}{2007}\natexlab{}.
\newblock \showarticletitle{The network structure of exploration and
  exploitation}.
\newblock \bibinfo{journal}{\emph{Administrative science quarterly}}
  \bibinfo{volume}{52}, \bibinfo{number}{4} (\bibinfo{year}{2007}),
  \bibinfo{pages}{667--694}.
\newblock


\bibitem[\protect\citeauthoryear{Levin and Levina}{Levin and Levina}{2019}]%
        {levin2019bootstrapping}
\bibfield{author}{\bibinfo{person}{Keith Levin} {and}
  \bibinfo{person}{Elizaveta Levina}.} \bibinfo{year}{2019}\natexlab{}.
\newblock \showarticletitle{Bootstrapping networks with latent space
  structure}.
\newblock \bibinfo{journal}{\emph{arXiv preprint arXiv:1907.10821}}
  (\bibinfo{year}{2019}).
\newblock


\bibitem[\protect\citeauthoryear{Lyzinski, Tang, Athreya, Park, and
  Priebe}{Lyzinski et~al\mbox{.}}{2016}]%
        {lyzinski2016community}
\bibfield{author}{\bibinfo{person}{Vince Lyzinski}, \bibinfo{person}{Minh
  Tang}, \bibinfo{person}{Avanti Athreya}, \bibinfo{person}{Youngser Park},
  {and} \bibinfo{person}{Carey~E Priebe}.} \bibinfo{year}{2016}\natexlab{}.
\newblock \showarticletitle{Community detection and classification in
  hierarchical stochastic blockmodels}.
\newblock \bibinfo{journal}{\emph{IEEE Transactions on Network Science and
  Engineering}} \bibinfo{volume}{4}, \bibinfo{number}{1}
  (\bibinfo{year}{2016}), \bibinfo{pages}{13--26}.
\newblock


\bibitem[\protect\citeauthoryear{March}{March}{1991}]%
        {march1991exploration}
\bibfield{author}{\bibinfo{person}{James~G March}.}
  \bibinfo{year}{1991}\natexlab{}.
\newblock \showarticletitle{Exploration and exploitation in organizational
  learning}.
\newblock \bibinfo{journal}{\emph{Organization Science}} \bibinfo{volume}{2},
  \bibinfo{number}{1} (\bibinfo{year}{1991}), \bibinfo{pages}{71--87}.
\newblock


\bibitem[\protect\citeauthoryear{Mengistu, Huizinga, Mouret, and
  Clune}{Mengistu et~al\mbox{.}}{2016}]%
        {mengistu2016evolutionary}
\bibfield{author}{\bibinfo{person}{Henok Mengistu}, \bibinfo{person}{Joost
  Huizinga}, \bibinfo{person}{Jean-Baptiste Mouret}, {and}
  \bibinfo{person}{Jeff Clune}.} \bibinfo{year}{2016}\natexlab{}.
\newblock \showarticletitle{The evolutionary origins of hierarchy}.
\newblock \bibinfo{journal}{\emph{PLoS computational biology}}
  \bibinfo{volume}{12}, \bibinfo{number}{6} (\bibinfo{year}{2016}),
  \bibinfo{pages}{e1004829}.
\newblock


\bibitem[\protect\citeauthoryear{Migliano, Battiston, Viguier, Page, Dyble,
  Schlaepfer, Smith, Astete, Ngales, Gomez-Gardenes, et~al\mbox{.}}{Migliano
  et~al\mbox{.}}{2020}]%
        {migliano2020hunter}
\bibfield{author}{\bibinfo{person}{Andrea~B Migliano},
  \bibinfo{person}{Federico Battiston}, \bibinfo{person}{Sylvain Viguier},
  \bibinfo{person}{Abigail~E Page}, \bibinfo{person}{Mark Dyble},
  \bibinfo{person}{Rodolph Schlaepfer}, \bibinfo{person}{Daniel Smith},
  \bibinfo{person}{Leonora Astete}, \bibinfo{person}{Marilyn Ngales},
  \bibinfo{person}{Jesus Gomez-Gardenes}, {et~al\mbox{.}}}
  \bibinfo{year}{2020}\natexlab{}.
\newblock \showarticletitle{Hunter-gatherer multilevel sociality accelerates
  cumulative cultural evolution}.
\newblock \bibinfo{journal}{\emph{Science advances}} \bibinfo{volume}{6},
  \bibinfo{number}{9} (\bibinfo{year}{2020}), \bibinfo{pages}{eaax5913}.
\newblock


\bibitem[\protect\citeauthoryear{Migliano, Page, G{\'o}mez-Garde{\~n}es,
  Salali, Viguier, Dyble, Thompson, Chaudhary, Smith, Strods,
  et~al\mbox{.}}{Migliano et~al\mbox{.}}{2017}]%
        {migliano2017characterization}
\bibfield{author}{\bibinfo{person}{Andrea~B Migliano},
  \bibinfo{person}{Abigail~E Page}, \bibinfo{person}{Jesus
  G{\'o}mez-Garde{\~n}es}, \bibinfo{person}{Gul~Deniz Salali},
  \bibinfo{person}{Sylvain Viguier}, \bibinfo{person}{Mark Dyble},
  \bibinfo{person}{James Thompson}, \bibinfo{person}{Nikhill Chaudhary},
  \bibinfo{person}{Daniel Smith}, \bibinfo{person}{Janis Strods},
  {et~al\mbox{.}}} \bibinfo{year}{2017}\natexlab{}.
\newblock \showarticletitle{Characterization of hunter-gatherer networks and
  implications for cumulative culture}.
\newblock \bibinfo{journal}{\emph{Nature Human Behaviour}} \bibinfo{volume}{1},
  \bibinfo{number}{2} (\bibinfo{year}{2017}), \bibinfo{pages}{1--6}.
\newblock


\bibitem[\protect\citeauthoryear{Newman}{Newman}{2006}]%
        {newman2006modularity}
\bibfield{author}{\bibinfo{person}{Mark~EJ Newman}.}
  \bibinfo{year}{2006}\natexlab{}.
\newblock \showarticletitle{Modularity and community structure in networks}.
\newblock \bibinfo{journal}{\emph{Proceedings of the national academy of
  sciences}} \bibinfo{volume}{103}, \bibinfo{number}{23}
  (\bibinfo{year}{2006}), \bibinfo{pages}{8577--8582}.
\newblock


\bibitem[\protect\citeauthoryear{Priebe, Park, Vogelstein, Conroy, Lyzinski,
  Tang, Athreya, Cape, and Bridgeford}{Priebe et~al\mbox{.}}{2019}]%
        {priebe2019two}
\bibfield{author}{\bibinfo{person}{Carey~E Priebe}, \bibinfo{person}{Youngser
  Park}, \bibinfo{person}{Joshua~T Vogelstein}, \bibinfo{person}{John~M
  Conroy}, \bibinfo{person}{Vince Lyzinski}, \bibinfo{person}{Minh Tang},
  \bibinfo{person}{Avanti Athreya}, \bibinfo{person}{Joshua Cape}, {and}
  \bibinfo{person}{Eric Bridgeford}.} \bibinfo{year}{2019}\natexlab{}.
\newblock \showarticletitle{On a two-truths phenomenon in spectral graph
  clustering}.
\newblock \bibinfo{journal}{\emph{Proceedings of the National Academy of
  Sciences}} \bibinfo{volume}{116}, \bibinfo{number}{13}
  (\bibinfo{year}{2019}), \bibinfo{pages}{5995--6000}.
\newblock


\bibitem[\protect\citeauthoryear{Rombach, Porter, Fowler, and Mucha}{Rombach
  et~al\mbox{.}}{2014}]%
        {rombach2014core}
\bibfield{author}{\bibinfo{person}{M~Puck Rombach}, \bibinfo{person}{Mason~A
  Porter}, \bibinfo{person}{James~H Fowler}, {and} \bibinfo{person}{Peter~J
  Mucha}.} \bibinfo{year}{2014}\natexlab{}.
\newblock \showarticletitle{Core-periphery structure in networks}.
\newblock \bibinfo{journal}{\emph{SIAM Journal on Applied mathematics}}
  \bibinfo{volume}{74}, \bibinfo{number}{1} (\bibinfo{year}{2014}),
  \bibinfo{pages}{167--190}.
\newblock


\bibitem[\protect\citeauthoryear{Rombach, Porter, Fowler, and Mucha}{Rombach
  et~al\mbox{.}}{2017}]%
        {rombach2017core}
\bibfield{author}{\bibinfo{person}{Puck Rombach}, \bibinfo{person}{Mason~A
  Porter}, \bibinfo{person}{James~H Fowler}, {and} \bibinfo{person}{Peter~J
  Mucha}.} \bibinfo{year}{2017}\natexlab{}.
\newblock \showarticletitle{Core-periphery structure in networks (revisited)}.
\newblock \bibinfo{journal}{\emph{SIAM review}} \bibinfo{volume}{59},
  \bibinfo{number}{3} (\bibinfo{year}{2017}), \bibinfo{pages}{619--646}.
\newblock


\bibitem[\protect\citeauthoryear{Smaldino, Moser, Velilla, and
  Werling}{Smaldino et~al\mbox{.}}{2022}]%
        {smaldino2022maintaining}
\bibfield{author}{\bibinfo{person}{Paul~E Smaldino}, \bibinfo{person}{Cody
  Moser}, \bibinfo{person}{Alejandro~P{\'e}rez Velilla}, {and}
  \bibinfo{person}{Mikkel Werling}.} \bibinfo{year}{2022}\natexlab{}.
\newblock \showarticletitle{Maintaining transient diversity is a general
  principle for improving collective problem solving}.
\newblock  (\bibinfo{year}{2022}).
\newblock


\bibitem[\protect\citeauthoryear{Sussman, Tang, Fishkind, and Priebe}{Sussman
  et~al\mbox{.}}{2012}]%
        {sussman2012consistent}
\bibfield{author}{\bibinfo{person}{Daniel~L Sussman}, \bibinfo{person}{Minh
  Tang}, \bibinfo{person}{Donniell~E Fishkind}, {and} \bibinfo{person}{Carey~E
  Priebe}.} \bibinfo{year}{2012}\natexlab{}.
\newblock \showarticletitle{A consistent adjacency spectral embedding for
  stochastic blockmodel graphs}.
\newblock \bibinfo{journal}{\emph{J. Amer. Statist. Assoc.}}
  \bibinfo{volume}{107}, \bibinfo{number}{499} (\bibinfo{year}{2012}),
  \bibinfo{pages}{1119--1128}.
\newblock


\bibitem[\protect\citeauthoryear{Sussman, Tang, and Priebe}{Sussman
  et~al\mbox{.}}{2013}]%
        {sussman2013consistent}
\bibfield{author}{\bibinfo{person}{Daniel~L Sussman}, \bibinfo{person}{Minh
  Tang}, {and} \bibinfo{person}{Carey~E Priebe}.}
  \bibinfo{year}{2013}\natexlab{}.
\newblock \showarticletitle{Consistent latent position estimation and vertex
  classification for random dot product graphs}.
\newblock \bibinfo{journal}{\emph{IEEE transactions on pattern analysis and
  machine intelligence}} \bibinfo{volume}{36}, \bibinfo{number}{1}
  (\bibinfo{year}{2013}), \bibinfo{pages}{48--57}.
\newblock


\bibitem[\protect\citeauthoryear{Tang and Priebe}{Tang and Priebe}{2018}]%
        {tang2018limit}
\bibfield{author}{\bibinfo{person}{Minh Tang} {and} \bibinfo{person}{Carey~E
  Priebe}.} \bibinfo{year}{2018}\natexlab{}.
\newblock \showarticletitle{Limit theorems for eigenvectors of the normalized
  Laplacian for random graphs}.
\newblock \bibinfo{journal}{\emph{The Annals of Statistics}}
  \bibinfo{volume}{46}, \bibinfo{number}{5} (\bibinfo{year}{2018}),
  \bibinfo{pages}{2360--2415}.
\newblock


\bibitem[\protect\citeauthoryear{Tudisco and Higham}{Tudisco and
  Higham}{2019}]%
        {tudisco2019nonlinear}
\bibfield{author}{\bibinfo{person}{Francesco Tudisco} {and}
  \bibinfo{person}{Desmond~J Higham}.} \bibinfo{year}{2019}\natexlab{}.
\newblock \showarticletitle{A nonlinear spectral method for core--periphery
  detection in networks}.
\newblock \bibinfo{journal}{\emph{SIAM Journal on Mathematics of Data Science}}
  \bibinfo{volume}{1}, \bibinfo{number}{2} (\bibinfo{year}{2019}),
  \bibinfo{pages}{269--292}.
\newblock


\bibitem[\protect\citeauthoryear{Udell and Townsend}{Udell and
  Townsend}{2019}]%
        {udell2019big}
\bibfield{author}{\bibinfo{person}{Madeleine Udell} {and} \bibinfo{person}{Alex
  Townsend}.} \bibinfo{year}{2019}\natexlab{}.
\newblock \showarticletitle{Why are big data matrices approximately low rank?}
\newblock \bibinfo{journal}{\emph{SIAM Journal on Mathematics of Data Science}}
  \bibinfo{volume}{1}, \bibinfo{number}{1} (\bibinfo{year}{2019}),
  \bibinfo{pages}{144--160}.
\newblock


\bibitem[\protect\citeauthoryear{Von~Luxburg}{Von~Luxburg}{2007}]%
        {von2007tutorial}
\bibfield{author}{\bibinfo{person}{Ulrike Von~Luxburg}.}
  \bibinfo{year}{2007}\natexlab{}.
\newblock \showarticletitle{A tutorial on spectral clustering}.
\newblock \bibinfo{journal}{\emph{Statistics and computing}}
  \bibinfo{volume}{17}, \bibinfo{number}{4} (\bibinfo{year}{2007}),
  \bibinfo{pages}{395--416}.
\newblock


\bibitem[\protect\citeauthoryear{Yanchenko and Sengupta}{Yanchenko and
  Sengupta}{2022}]%
        {yanchenko2022core}
\bibfield{author}{\bibinfo{person}{Eric Yanchenko} {and}
  \bibinfo{person}{Srijan Sengupta}.} \bibinfo{year}{2022}\natexlab{}.
\newblock \showarticletitle{Core-periphery structure in networks: a statistical
  exposition}.
\newblock \bibinfo{journal}{\emph{arXiv preprint arXiv:2202.04455}}
  (\bibinfo{year}{2022}).
\newblock


\bibitem[\protect\citeauthoryear{Zhu and Ghodsi}{Zhu and Ghodsi}{2006}]%
        {zhu2006automatic}
\bibfield{author}{\bibinfo{person}{Mu Zhu} {and} \bibinfo{person}{Ali Ghodsi}.}
  \bibinfo{year}{2006}\natexlab{}.
\newblock \showarticletitle{Automatic dimensionality selection from the scree
  plot via the use of profile likelihood}.
\newblock \bibinfo{journal}{\emph{Computational Statistics \& Data Analysis}}
  \bibinfo{volume}{51}, \bibinfo{number}{2} (\bibinfo{year}{2006}),
  \bibinfo{pages}{918--930}.
\newblock


\end{thebibliography}

%%%%%%%%%%%%%%%%%%%%%%%%%%%%%%%%%%%%%%%%%%%%%%%%%%%%%%%%%%%%%%%%%%%%%%%%

\end{document}